\documentclass[sigconf]{acmart}

\usepackage{booktabs} 
\usepackage{ifthen}
\usepackage{graphicx}
\usepackage{subcaption}
\usepackage{enumitem}
\usepackage{multirow}

\newif\ifcomment
\commenttrue 

\ifcomment

\else  
\fi

\definecolor{heraldBlue}{rgb}{0.0,0.0,0.8}
\definecolor{heraldRed}{rgb}{0.8,0.0,0.0}
\definecolor{heraldGray}{rgb}{0.4,0.4,0.4}
\definecolor{heraldBlack}{rgb}{0.0,0.0,0.0}
\definecolor{heraldGreen}{rgb}{0.0,0.4,0.0}

\newcommand{\descr}[1]{\smallskip\noindent\textbf{#1}}

\setcopyright{none}
\renewcommand\footnotetextcopyrightpermission[1]{}

\begin{document}
\title{User Identity Linkage in Social Media Using Linguistic and Social Interaction Features}

\author{Despoina Chatzakou$^{\dagger}$, Juan Soler-Company$^{\ddagger}$, Theodora Tsikrika$^{\dagger}$ \and Leo Wanner$^{\ddagger\clubsuit}$, Stefanos Vrochidis$^{\dagger}$, Ioannis Kompatsiaris$^{\dagger}$} 
\affiliation{
\vspace{0.5cm}
$^\dagger$Information Technologies Institute, Centre for Research and Technology Hellas,
$^\ddagger$Pompeu Fabra University, $^\clubsuit$ICREA\\
dchatzakou@iti.gr, juan.soler@upf.edu, theodora.tsikrika@iti.gr\\
leo.wanner@upf.edu, stefanos@iti.gr, ikom@iti.gr
\country{} }

\renewcommand{\shortauthors}{D. Chatzakou et al.}

\begin{abstract}
Social media users often hold several accounts in their effort to multiply the spread of their thoughts, ideas, and viewpoints.
In the particular case of objectionable content, users tend to create multiple accounts to bypass the combating measures enforced by social media platforms and thus retain their online identity even if some of their accounts are suspended. 
User identity linkage aims to reveal social media accounts likely to belong to the same natural person so as to prevent the spread of abusive/illegal activities. To this end, this work proposes a machine learning-based detection model, which uses multiple attributes of users' online activity in order to identify whether two or more virtual identities belong to the same real natural person.
The models efficacy is demonstrated on two cases on abusive and terrorism-related Twitter content.
\end{abstract}

\keywords{Actor identity resolution, Abusive and Illegal content, Twitter}

\maketitle

\section{Introduction}\label{sec:intro}

In its somewhat more than 20 years of existence, social media have become an integral part of the life of more than $2.6B$ people around the globe. 
Originally envisaged as a means to
stay connected with friends, get informed, or be entertained, it has become a very powerful instrument for public opinion formation and dissemination of all kinds of not always harmless content.
Particularly worrying is the spread of abusive, extremist, and terrorism-related content via widely used online social platforms, such as Twitter and Facebook. 
In order to address this problem, social media administrators implement filtering methods and suspend accounts once harmful content is detected~\cite{twitterCompanySuspension}. 

However, to counter such measures and overcome the suspension policies, users seeking to widely disseminate deleterious material often follow various strategies, the most popular being the setting up of multiple (back-up) accounts that allow them to keep contact with individuals with the same disposition (e.g., violent extremists) 
and exchange content, even after one of their accounts gets suspended~\cite{fisher2015swarmcast, klausen2015tweeting}.
It is thus of paramount importance to be able to detect user accounts (alias \textit{user identities}) likely to belong to the same  person, so as to stop the propagation of harmful behavior on a large scale, including the spread of abusive or terrorism-related material.\footnote{Linking users is also important in other contexts, e.g., curbing the spread of spam or fake content.}

User identity linkage (i.e., detection of multiple user identities) has been studied both \textit{across} social networks (e.g., \cite{liu2014hydra,riederer2016linking}) and \textit{within} the same social network (e.g., \cite{tsikerdekis2014multiple,johansson2015timeprints}). 
This paper focuses on the latter case and, particularly, on Twitter.
Twitter has been selected as it is one of the most popular social media platforms and often contains abusive~\cite{badjatiya2017deep,chatzakou2017mean} or terrorism-related~\cite{conway2018disrupting,gialampoukidis2017detection} material. 
Moreover, Twitter is a rather challenging platform for investigating this phenomenon, since tweets are short and often contain grammatical and orthographic errors, thus making it harder to use off-the-shelf natural language processing tools to analyze them in the context of such investigations. 
As a consequence, Twitter is often avoided as a single social media source for the study of user identity linkage.
Furthermore, user identity linkage research has thus far been mainly conducted on English data sources. 
Since the dissemination of deleterious (e.g., abusive and terrorism-related) material is not limited to English, the consideration of other languages is also necessary.

\descr{Overview \& Contributions.}
In this paper, we design, implement, and evaluate a methodology geared to identify the linkage between online user accounts within the same social network.
Specifically, this work proposes a framework which considers a wide range of profile, linguistic, activity, and network characteristics (the latter two are also referred to as \emph{social interaction} features) for representing users' online presence, and employs machine learning and deep learning-based classifiers for identifying accounts potentially linked to the same natural person.
Our main contributions can be summarized as follows: to the best of our knowledge, this is the first user identity linkage work to employ (i) a wide range of features extracted from social networks constructed based on users' activity, (ii) advanced syntactic features based on dependency trees, (iii) semantic similarities based on word embeddings, and (iv) deep neural networks in such a classification setup.
Moreover, comprehensive evaluation experiments are performed on two Twitter datasets related to abusive behaviors and terrorism phenomena, with English and Arabic material, respectively, and the experimental results are promising, achieving up to $99.50\%$ AUC.

The rest of the paper is organized as follows. 
Section~\ref{sec:relatedwork} reviews the related work.
Section~\ref{sec:methodology} presents the proposed framework, the extracted features, and the techniques for modeling the data, and predicting possible user linkage.
Section~\ref{sec:experiments} describes the employed datasets, the process for constructing the ground truth, and the experimental methodology, while Section~\ref{sec:classificationResults} presents the experimental results.
Finally, Section~\ref{sec:conclusions} draws some conclusions and outlines future work.

\section{Related  Work}\label{sec:relatedwork}

Numerous studies have examined user identity linkage \textit{across} online social networks; see, e.g.,~\cite{malhotra2012studying,liu2014hydra,riederer2016linking}.
Malhotra et al.~\cite{malhotra2012studying} proposed to disambiguate profiles of the same user based on their digital footprint in both Twitter and LinkedIn.
Twitter has also been jointly considered in many works
as one of the studied platforms 
in relation to other social networks, e.g., Yelp~\cite{goga2013exploiting}, Flickr~\cite{goga2013exploiting}, 
Foursquare~\cite{riederer2016linking}, Instagram~\cite{riederer2016linking}, and Facebook~\cite{liu2014hydra}.
For instance, authors in~\cite{riederer2016linking} proposed a method that examines whether two accounts belong to the same mobile user by exploiting location information, when they are active on both Twitter and Instagram.

Identity linkage \textit{within} a single social network has also been explored.
For instance, an Irish forum was studied~\cite{johansson2015timeprints} to first unmask authors identities and then detect matching aliases.
The so-called `sockpuppetry' (i.e., blocked users initiating new accounts) has been considerably studied on Wikipedia~\cite{solorio2013case,tsikerdekis2014multiple}.
Finally, user identity linkage has been explored on popular 
online news sites, such as \textit{The Guardian} and the \textit{SPIEGEL ONLINE}, 
to assist their providers detect manipulations of public opinion~\cite{pennekamp2019hi}. 

Profile, content, and network attributes are often exploited to build such detection models. 
User name, screen name, and biography are common profile attributes~\cite{goga2015reliability,mu2016user}.
In relation to the posted content, temporal (e.g., timestamps) and spatial (e.g., geotags) information \cite{johansson2015timeprints,riederer2016linking,nie2016identifying},
as well as stylometric features (e.g., part-of-speech n-grams, etc.)~\cite{solorio2013case,johansson2015timeprints,pennekamp2019hi}
are widely employed.
The way that a user's social network is formulated and their communication patterns can also provide useful information about a user's identity; hence, network attributes have been used to detect actor's identity across multiple social networks~\cite{nie2016identifying,liu2016aligning}.
For instance, a user's immediate or non-immediate neighborhood can be exploited by considering friendship relations.

Building upon such features, supervised, unsupervised, and semi-supervised methods have been considered.
For instance, a probabilistic classification based on Naive Bayes has been employed to link user identities across social media~\cite{zafarani2013connecting}.
Decision Trees, SVM, and kNN algorithms have also been tested~\cite{malhotra2012studying}.
Moreover, an alignment algorithm has also been used, where an affinity score based on timestamped sparse and dense location-based properties is computed to find the most likely matching identities using a maximum weighted matching scheme~\cite{riederer2016linking}.
Regarding semi-supervised models, a multi-objective framework has been built for modeling heterogeneous behaviors and structural consistency maximization~\cite{liu2014hydra}.

\begin{table}[!t]
\caption{Comparison of our method with past works. A: Activity, L: Linguistic (CH: character, W: word, S: sentence, D: dictionary, SY: syntactic), N: Network (DI: distribution, SE: segmentation, CO: connection).}
\label{tbl:relatedAlternatives}
\scalebox{0.78}{
\begin{tabular}{|l|c|c|c|c|c|c|c|l|l|c|c|}
\hline
                                                                & \multicolumn{9}{c|}{\textbf{Features}}                                                                                                                                                                                                                                                           & \multicolumn{2}{c|}{\textbf{ML method used}}                                                                                \\ \hline
\textbf{\begin{tabular}[c]{@{}l@{}}Related\\ Work\end{tabular}} & \multicolumn{1}{c|}{\textbf{A}} & \multicolumn{5}{c|}{\textbf{L}}                                                                                                                                           & \multicolumn{3}{c|}{\textbf{N}}                                                    & \multicolumn{1}{c|}{\textbf{Classic}} & \multicolumn{1}{c|}{\textbf{\begin{tabular}[c]{@{}l@{}}Neural\\ Nets\end{tabular}}} \\ \hline
                                                                & \multicolumn{1}{c|}{}           & \multicolumn{1}{c|}{\textbf{CH}} & \multicolumn{1}{c|}{\textbf{W}} & \multicolumn{1}{c|}{\textbf{S}} & \multicolumn{1}{c|}{\textbf{D}} & \multicolumn{1}{c|}{\textbf{SY}} & \multicolumn{1}{c|}{\textbf{DI}} & \textbf{SE}            & \textbf{CO}            & \multicolumn{1}{c|}{}                 & \multicolumn{1}{c|}{}                                                               \\ \hline
\cite{johansson2015timeprints}                 & x                               & x                                & x                               &                                 & x                               &                                  &                                  &                        &                        & x                                     &                                                                                     \\ \hline
\cite{johansson2013detecting}                  & x                               & x                                & x                               &                                 & x                               &                                  &                                  &                        &                        &                                       &                                                                                     \\ \hline
\cite{solorio2013case}                         &                                 & x                                & x                               & x                               &                                 & x                                &                                  &                        &                        & x                                     &                                                                                     \\ \hline
\cite{tsikerdekis2014multiple}                 & x                               &                                  &                                 &                                 &                                 &                                  &                                  &                        &                        & x                                     &                                                                                     \\ \hline
\cite{kumar2017army}                           & x                               & x                                & x                               & x                               & x                               & x                                &                                  & \multicolumn{1}{c|}{x} & \multicolumn{1}{c|}{x} & x                                     &                                                                                     \\ \hline
\cite{pennekamp2019hi}                         & x                               & x                                & x                               & x                               &                                 & x                                &                                  &                        &                        &                                       &                                                                                     \\ \hline
\textbf{This work}                                              & x                               & x                                & x                               & x                               & x                               & x                                & x                                & \multicolumn{1}{c|}{x} & \multicolumn{1}{c|}{x} & x                                     & x                                                                                   \\ \hline
\end{tabular}}
\end{table}

Table~\ref{tbl:relatedAlternatives} compares our method to those that are most relevant to our problem setting (i.e., identity linkage \textit{within} the same platform).
Most of such works use ``classic'' (traditional) machine learning classifiers, such as SVMs~\cite{johansson2015timeprints,solorio2013case,tsikerdekis2014multiple}, Naive Bayes~\cite{johansson2015timeprints}, and Random Forest~\cite{tsikerdekis2014multiple,kumar2017army}. 
Moreover, matching approaches based on similarity measures (e.g., cosine similarity or euclidean distance)~\cite{johansson2013detecting}, as well as threshold-based approaches have also been employed~\cite{pennekamp2019hi}.
Under the features category three main types of features are listed, i.e., activity-, linguistic-, and network-based.
Depending on the considered platform, different activity-based features are used, such as number of posts and replies, down- and up-votes, number of total revisions, etc.
Moreover, users' activity is often examined in relation to the temporal dimension, by considering for instance the mean time between two consecutive posts or the posting activity in relation to different timeframes (such as hours, period of day, and month).
The linguistic-based features are highly related to a user's behavioral and writing style, as for instance average words length, average number of characters per word and/or sentence, upper-cased letters, and part-of-speech tags (such as verbs, nouns, and adverbs).
Finally, the network-based features so far have been related to a reply-based network~\cite{kumar2017army}, examining users' tendency to cluster with others (based on clustering coefficient) and quantifying the extent to which users reciprocate the reply communication they receive from other users (reciprocity).
Overall, apart from English, Irish~\cite{johansson2015timeprints,johansson2013detecting} and German~\cite{pennekamp2019hi} textual sources have been studied.

\descr{Contributions.} 
Compared to existing works, we use a wide range of linguistic features (driven by well-established approaches used in similar tasks, e.g., author profiling and identification), while to our knowledge we are the first to employ dependency and tree features in addition to part-of-speech (as syntactic features) in this context.
Moreover, we advance state-of-the-art by considering various social interaction features, which contribute significantly in successfully detecting accounts likely to belong to the same person within a social network.
Specifically, we employ a ``conversation-based network'', which considers mentions, replies, and retweets, to first construct the network and estimate then various network features. 
To the best of our knowledge, we are the first to employ the conversation-based network and all these features in this context.

To be in alignment with the literature, we evaluate various traditional machine learning methods, i.e., probabilistic, tree-based, and ensemble classifiers.
In addition, we study the application of deep learning on the user identity linkage task.
The designed neural network architecture digests both textual information and various numerical metadata (i.e., activity, linguistic, and network features).
Finally, since the propagation of objectionable material is not limited to English, we conduct comprehensive experiments in two case studies related to abusive and terrorism phenomena, associated with English and Arabic textual sources, respectively.

\section{Discovery of Account Linkage}\label{sec:methodology}

This section details the proposed framework for detecting the possible linkage of user accounts
in social media based on models of user behavior.
To this end, a wide range of user characteristics are considered for representing users' online presence, and, based on these extracted features, machine learning and deep learning-based classifiers are employed for distinguishing between \textit{linked accounts} (i.e., accounts belonging to the same person) and non-linked accounts.

\subsection{Individual User Account Features}\label{subsec:featuresextraction}

Various attributes can be exploited in social media to model the behavior of \textit{each individual user}, namely:
\begin{enumerate}
\item \textit{Profile Features} (P) extracted from a user's profile, such as demographic information, biography, avatar (i.e., image provided by the user to visually present themselves), etc.
\item \textit{Activity Features} (A) related to a user's posting behavior, such as number of posts, replies, mentions, etc.
\item \textit{Linguistic Features} (L) extracted from  users' posted content that may be used to model users with respect to, e.g., their writing style or topics of interest.
\item \textit{Network Features} (N) extracted from the social network interactions between users.
\end{enumerate}
Below, we detail the set of features considered per \textit{individual user account} for each of the aforementioned categories.

\descr{Profile Features.}
Features in this category include the age of the account (i.e., number of days since its creation), whether the account is verified or not (i.e., acknowledged by Twitter as an account linked to a user of ``public interest'), and whether or not the user has provided information about their location.

\descr{Activity Features.}
These features provide an overview of a user's online presence with respect to the considered social network and include the number of: posts, lists subscribed to, shares, favorited tweets, mentions, and hashtags, as well as the posts' inter-arrival time.
For instance, mentions can be used to directly interact with another user (and possibly perform direct attacks in an abusive context), while the use of hashtags (particularly of popular ones) is a way to increase a post's visibility.

\descr{Linguistic Features.}
This set of features analyzes the writing style of the author of a tweet. 
Based on the posted content, surface-oriented and deeper stylistic features are extracted.
In particular, five subcategories of features are considered~\cite{soler2017relevance}, as described next.

1. \emph{Character-based features:} ratio of the number of each of the following characters to the total number of characters: upper-cased, periods, commas, parentheses, exclamations, colons, number digits, semicolons, hyphens, and quotation marks.

2. \emph{Word-based features:} mean number of characters per word, vocabulary richness (i.e., different words being used), acronyms, stopwords, first person pronouns, usage of words composed by two or three characters, standard deviation (STD) of word length, and the difference between the longest and shortest words.

3. \emph{Sentence-based features:} mean number and standard deviation of words per sentence, and difference between the maximum and minimum number of words per sentence in a text.

4. \emph{Dictionary-based features:} the ratio of each of the following types of tokens to the total number of words in a text: discourse markers, interjections, abbreviations, curse words, and polar (positive/negative) words~\cite{hu2004mining}.

5. \emph{Syntactic features:} three types of syntactic features are taken into account: (i) Part-of-Speech (POS) features: relative frequency of each POS tag in a text; (ii) Dependency features: occurrence of syntactic dependency relations in the dependency trees of the text;\footnote{Syntactic dependency trees are unordered rooted trees that represent the syntactic structure of a sentence according to a specific grammar. Their nodes correspond to the words of the sentence and are connected via binary asymmetrical dependencies.} to this end, we extract the frequency of each individual dependency relation per sentence, the usage ratio of the passive voice, and the number of coordinate/subordinate clauses per sentence; and (iii)  Tree features: measures of  the tree width, the tree depth, and the ramification factor, where \textit{tree depth} is defined as the maximum number of nodes between the root and a leaf node, \textit{tree width} is the maximum number of siblings at any of levels of the tree, and the \textit{ramification factor} is the mean number of children per level; in other words, the tree  features characterize the complexity of the inner structure of the sentences (simple clauses, as well as subordinate and coordinate clauses).
To extract syntactic features, the parser presented in~\cite{mcdonald2006multilingual} has been trained on English and  Arabic material annotated with Universal Dependencies.

\descr{Network Features.}
This feature category aims to measure the popularity of a user based on different criteria, such as the number of followers ({\it in-degree centrality}), friends ({\it out-degree centrality}), and their ratio; since Twitter allows users to follow anyone without their approval, this ratio can quantify a user's popularity.
Overall, these measures can quantify a user's opportunity to have a positive or negative impact in their ego-network in a direct way.

To dig deeper into users' relations, we construct a ``conversation-based network'' based on the mentions, replies, and retweets between each pair of users,
and extract (using Gephi~\cite{gephi}) six network features grouped as follows: (i) Distribution metrics: hub, authority, Eigenvector, and PageRank centralities, which measure users' influence and connectivity in their immediate and extended neighborhoods, (ii) Connection metric: number of triangles a node belongs to, and (iii) Segmentation metric: Clustering Coefficient, which shows a user's tendency to cluster with others.
To the best of our knowledge, we are the first to employ the conversation-based network and all these features in this context.

\subsection{User Modeling}\label{subsec:usermodeling}

The aforementioned feature categories (or \textit{sets}) $S$$=$$\{P, A, L, N\}$ can be exploited to model the behavior of each \textit{individual user account} in a social media platform.
We thus define the feature vector for each user $u_i$ and feature category $S$ as ${V_S}_{u_i}$$=$$<{f_S}_{i_1}, {f_S}_{i_2}, \ldots,  {f_S}_{i_n}>$,
where ${f_S}_{i_j}$ is the $j$th feature of category $S$ for user $u_i$, and $n$ equals to the total number of included features for this category.
For instance, for the network features category, a feature vector can be created for every $u_i$ as follows: ${V_N}_{u_i}$=$ 
<authority_i, triangles_i, eigenvector_i,$ $pagerank_i, coef_i, hub_i>$.
A feature vector $V_{{All}_{u_i}}$ can also be created by considering all features from all four sets.

To detect whether two accounts are likely to belong to the same person, we also need to jointly represent each \textit{user pair} so as to determine their potential relationship and use that as input to the classifier.
To this end, we jointly represent the behavior of each pair of users $u_i$ and $u_j$, $\forall$$i,j$, where $i$ $\neq$ $j$, as either (i) a feature vector of the absolute differences between the individual feature vectors of $u_i$ and $u_j$, or (ii) as a vector of four similarity scores, each estimated based on the similarity of the per-category $\{P, A, L, N\}$ feature vector.
To estimate these similarities, the cosine similarity, the Euclidean, and the Manhattan distance are used; for the latter two, normalization is applied, such that values $\in [0,1]$.

Apart from the above approaches to user pair modeling that take into account the extracted features, we can also measure the direct similarity of the evidence associated with each user, such as their posted content, social network, and profile. 
In particular, we focus on the similarity between the posts of two users, since users tend to express themselves in standard ways by frequently using the same words or expressions; moreover, due to daily social interactions, even different persons may result in using the same words in essentially the same way~\cite{babcock2014latent}.
We thus consider two additional features corresponding to the similarities between the  posts of two users, measured in terms of their (i) \textit{edit distance}, i.e., number of changes needed to convert a text to another,
and (ii) \textit{semantic similarity}. To this end, a preprocessing step is applied to remove all numbers, mentions, and URLs from the posts.

\emph{Edit distance} is estimated with the Levenshtein distance~\cite{Navarro2001ApproximateStringMatching}, which counts the minimum number of single-character edits needed to convert one string into another; for each pair of users, this is averaged out over all pairs of their posts.
\emph{Semantic similarity} is estimated based on a vector space model approach, whereby each word in a post is represented as a word embeddings vector.
Word embeddings allow modeling both semantic and syntactic relations of words, thus capturing more refined attributes and contextual cues inherent in language.
Specifically, we use Word2Vec~\cite{MikolovWordEmbedding2013} to: (1)~first establish a vocabulary based on the words included in the set more times than a user-defined threshold, (2)~apply a learning model so as to learn the words' vector representations in a $D$-dimensional space ($50$-$300$ dimensions can model hundreds of millions of words with high accuracy~\cite{MikolovWordEmbedding2013}), and (3)~output a vector representation for each word encountered in the input texts.
Based on~\cite{MikolovWordEmbedding2013} $50$-$300$ dimensions can model hundreds of millions of words with high accuracy.
Given the vector representations of all words in a post, the overall vector representation of the post is derived by averaging the vectors of all its words.
Finally, the set of all posts by a user, referred to as document $d$, is represented as a vector which contains the semantic center of all posts' vectors, $p$: $Sem_{center} (d) = \sum_{p \in d} vec(p) / |d|$, where $|d|$ is the number of the user's posts.

\subsection{Classification}\label{subsec:classification}

To be in alignment with the state-of-the-art, here, we proceed with both traditional machine learning methods and deep neural networks (NNs).
Regarding the former, probabilistic (e.g., Naive Bayes, BayesNet), tree-based (e.g., J48, LADTree, LMT), and ensemble classifiers are considered.
As an ensemble classifier, we use Random Forest which constructs a forest of decision trees with random subsets of features during classification; an important advantage is its ability to reduce overfitting by averaging several trees during model construction.
Moreover, Random Forests are quite efficient in terms of the time needed to train a model.
To build the Random Forest classifier, we tune the number of generated trees to $100$, while there is no limit set to the maximum depth.

\begin{figure}[!t]
	\centering
	\includegraphics[width=0.48\textwidth]{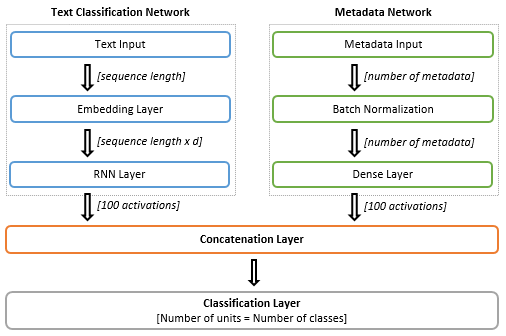}
	\caption{Deep neural network setup.}
	\label{fig:nn_setup}
\end{figure}

Even though the traditional machine learning approaches have been extensively used in similar tasks, they face an important drawback:
they cannot successfully combine semantic and cultural nuances of the written language.
For instance, taking into account the negation of words or sarcastic expressions with traditional machine learning approaches is a quite challenging task, as the structure of the sentence has to be effectively
presented in the set of features.
To overcome such difficulties, deep learning algorithms have been proposed that build upon neural networks. 
Therefore, here we also proceed with a modeling process building upon neural networks.
Specifically, in the neural network setup, we build a model to combine raw text with metadata (i.e., profile, activity, linguistic, network, and user pair features), similar to~\cite{founta2018unified}.
The combination of raw text with additional behavioral facts (such as users' popularity, social network, and account settings) allows us to capture different facets of users' behavior, and thus possibly detecting more efficiently accounts likely to belong to the same user.
Specifically, we construct a single network architecture which combines both text classification and metadata networks (see below) before their inputs are translated into classification probabilities.
Figure~\ref{fig:nn_setup} depicts the deep neural network setup used in this work.

\descr{Text Classification Network.}
We employ a Recurrent Neural Network (RNN)~\cite{mikolov2010recurrent}, which processes sequential data using recurrent connections between their neural activations at consecutive time steps.
RNNs were selected over other NN models since they have proven successful in understanding word sequences and interpreting their meaning. 
Specifically, we build upon a Gated Recurrent Unit (GRU) since it performs well on short texts (such as tweets)~\cite{founta2018unified}.
We employ a GRU with 100 units (neurons); we experimented with different sizes and this gave the best results for both datasets.
To avoid over-fitting, we use a recurrent dropout with $p = 0.5$.
Before moving through the RNN layers, the first layer performs a word embedding lookup, where all words are represented as high-dimensional vectors.
For English, we use pre-trained word vectors from Twitter~\cite{pennington2014glove}; for Arabic, we use AraVec~\cite{soliman2017aravec}, a pre-trained distributed word representation.
Tweets' words are mapped onto $200$ and $300$ dimensional vectors, for English and Arabic, respectively.

\descr{Metadata Network.}
After feeding the data to the metadata neural network, a batch normalization layer is used to enable faster learning and higher overall accuracy.
To learn the metadata, we use a simple dense layer with $100$ units, i.e., the same dimensionality as the text classification network.
Finally, we use \textit{tanh} as activation function, since it performs well with standardized numerical data.

\descr{Combined Network.}
We combine the text classification and metadata networks using a concatenation layer using a fully connected output layer (i.e., dense layer) with one neuron per class we want to predict and {\it softmax} as activation function.

\section{Experiments}\label{sec:experiments}

This section presents our evaluation experiments on abusive and terrorism-related datasets collected from Twitter.

\subsection{Datasets}
The first step is to collect the necessary content from Twitter, i.e., one of the most popular social networks with $\sim$$330M$ monthly active users~\cite{twitterStats}, which also gives access to an important number of sample tweets via its open API.
For our study, two datasets obtained from Twitter are used; we focus on these datasets since they are likely to involve users with multiple accounts~\cite{fisher2015swarmcast,klausen2015tweeting,cnetfake}.
It should be noted that the collected data correspond to publicly available data, we did not attempt to de-anonymize users, and we fully comply with the terms of use of the APIs we use.

\descr{Abusive Dataset.}
The dataset provided by~\cite{chatzakou2017measuring} was used for studying abusive activities on Twitter.
The authors collected a set of tweets between June and August 2016, using snowball sampling around the GamerGate controversy~\cite{Massanari09102015}, which is known to have produced many instances of cyber-bullying and cyber-aggression.
GamerGate originated from alleged improprieties in video game journalism, which quickly grew into a larger campaign centered around sexism and social justice.
The GamerGate controversy, and more specifically the hashtag \#GamerGate, can serve as a relatively unambiguous reference to posts that are likely to involve abusive/aggressive behavior from a fairly mature and hateful online community, since individuals on both sides of the controversy were using this hashtag. Moreover, extreme cases of bullying and aggressive behavior (e.g., direct threats of rape and murder) have been associated with it.
Overall, the dataset consists of $600k$ tweets in English and $312k$ users.

\descr{Terrorism Dataset.}
This dataset was created using Twitter's Search API, which returns tweets matching specified keywords.
Specifically, we collected data from February 2017 to June 2018 using a set of terrorism-related Arabic keywords provided by Law Enforcement and domain experts.
The dataset consists of $65k$ tweets and $35k$ users.
Based on a language detection library~\cite{nakatanilangdetect}, $99\%$ of the posts in our dataset are in Arabic.

\subsection{Ground Truth}\label{subsec:groundtruth}
Due to the absence of ground truth that indicates which user accounts belong to the same person, the ground truth for each dataset is created as follows.
First, we filter out all users with less than 10 posts (thus removing all users associated with insufficient evidence), and then we randomly select a subset of user accounts (e.g., $X$$=$$200$ users) by applying a stratified random sampling.
To this end, the entire population is first divided into homogeneous groups based on the number of posted tweets; this number is varied between $10$ and $60$ with step $5$, while the final group contains all users with more than $60$ posts. Then a random sample is selected from each group, with the sample size being proportional to the group's size compared to the entire population.

As in~\cite{johansson2015timeprints,pennekamp2019hi}, where no annotated datasets were available, 
we build the ground truth 
by splitting the posts of each selected user into two subsets, assigning to each subset a different user id (e.g., user $u_i$ becomes $u_{ia}$ and $u_{ib}$, and the tweets of $u_i$ are split between $u_{ia}$ and $u_{ib}$).
Thus, we come up with a dataset with the double number of user accounts (e.g., $400$ users for $X$$=$$200$) and a set of known \textit{linked accounts} (i.e., accounts belonging to the same person).
Two approaches are considered for splitting the tweets of the original accounts (e.g., $u_i$) into linked users (e.g., $u_{ia}$ and $u_{ib}$): (i) \textit{random assignment} of an equal number of posts to each, and (ii) \textit{interleaving}, where posts are initially sorted based on their timestamps and then alternately assigned to each of the linked accounts.

Hence, we have  two sets of users available: $A$$=$$ \{u_{1a}, u_{2a}, \ldots, u_{Xa}\}$ and $B$$=$$\{u_{1b}, u_{2b}, \ldots, u_{Xb}\}$.
Comparing each user $u_{ia}$ from set A with each user $u_{jb}$ in set B  $\forall$$i,j$, where $i$ $\neq$ $j$, we result to overall $Y = X * (X-1)$ user pairs (e.g., for $X = 200$, $Y = 39,800$), with each user pair in $Y$ corresponding to a non-linked account. 
For each dataset, we opt for maintaining a proportion of $10\%$ of linked and $90\%$ of non-linked accounts, given that previous works, e.g.,~\cite{kayes2015ya-abuse}, have indicated that about $10\%$ of  users within a dataset tend to exhibit bad behavior.
Therefore, for a given $X$, we randomly sample from $Y$ so as to reflect the above observation; e.g, for $X$$=$$200$, the final dataset contains $200$ linked accounts ($u_{ia}$, $u_{ib}$) and $Z = 9 \times 200 = 1,800$ non-linked accounts ($u_{ia}$, $u_{jb}$), $i$ $\neq$ $j$.  

We also (i) vary the number of randomly selected users $X$ from $200$ to $500$ in steps of $100$, and (ii) create unbalanced datasets by increasing the non-linked accounts; for this, we keep the same number of linked accounts and incrementally increase the number of non-linked accounts with step $9 \times X$.
E.g., for $X$$=$$200$, $Z$ ranges from $1,800$ to $39,800$ with step $9 \times 200 = 1,800$.
In the last step, we consider all $39,800$ (rather than the $39,600$) non-linked accounts.

\subsection{Features Selection}\label{subsec:featuresSelection}

\begin{table}[!t]
\centering
\caption{Individual user account features considered.}
\label{tbl:used_features}
\scalebox{0.71}{
\begin{tabular}{l|lp{7.2cm}l}
\hline
\textbf{Category}        & \multicolumn{2}{l}{\textbf{Features}}                                                                                                                                                                                                                                               \\ \hline
Activity                 & \multicolumn{2}{l}{avg. \# mentions, avg. \# hashtags, posts' inter-arrival time}                                                                                                                                                                                                    \\ \hline\hline
\multirow{1}{*}{Linguistic} & \textit{Character-based}  & ratios of upper-cased characters, periods, commas, parentheses, exclamations, colons, number digits, semicolons, hyphens and quotation marks, w.r.t. \# characters in a text                                                                    \\ \cline{2-3} 
                         & \textit{Word-based}       & mean \# characters per word, vocabulary richness, acronyms, stopwords, first person pronouns, usage of words composed by 2 or 3 characters, STD of word length, difference between the longest and shortest words            \\ \cline{2-3} 
                         & \textit{Sentence-based}   & mean and STD of words per sentence, difference between the max. and min. number of words per sentence                                                                                                   \\ \cline{2-3} 
                         & \textit{Dictionary-based} & ratios of discourse markers, interjections, abbreviations, curse words, polar words w.r.t. the \# of words in a text \\ \cline{2-3} 
                         & \textit{Syntactic-based}        & part-of-speech, dependency features, tree features                                                                                                                                                                                                                \\  \hline\hline
Network                  & \multicolumn{2}{l}{authority, hub, \# triangles, eigenvector, pagerank, clustering coefficient}                                                                                                                                                               \\ \hline
\end{tabular}}
\end{table}

Section~\ref{subsec:featuresextraction} described various features that could be considered for exploring whether two accounts belong to the same person.
Given the ground truth creation process applied in this work, profile features, as well as the number of followers, friends, and their ratio of the network features are excluded (as they would be the same for both linked accounts),
while for activity features (for the same reason) we can only consider the number of mentions and hashtags, and the posts' inter-arrival time.
Table~\ref{tbl:used_features} summarizes the examined features; in real scenarios, all features from the four categories could be considered and may be beneficial for the classification.

As expected, some of the features in Table~\ref{tbl:used_features} could be more distinguishing and thus assist more the classification.
To this end and towards feature selection, we examine the significance of differences between the distributions of linked and non-linked user accounts based on the two-sample Kolmogorov-Smirnov test.
This test is used since it enables to assess whether two samples come from the same distribution based on their empirical distribution function (ECDF). 
We consider as statistically significant all cases with $p$$<$$0.01$.
Due to space limits, we only present the ECDF plots of some features; to improve readability, some plots are trimmed.

\begin{figure}[t]
	\centering
	\begin{subfigure}[b]{0.23\textwidth}
			\captionsetup{font=scriptsize}
			\includegraphics[width=\textwidth]{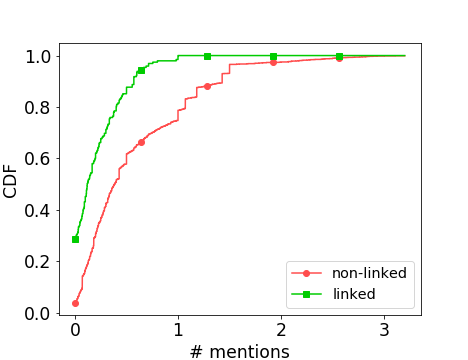}
			\caption{Mentions.}
			\label{fig:ecdf-en-mentions}
	\end{subfigure}
	\begin{subfigure}[b]{0.23\textwidth}
			\captionsetup{font=scriptsize}
			\includegraphics[width=\textwidth]{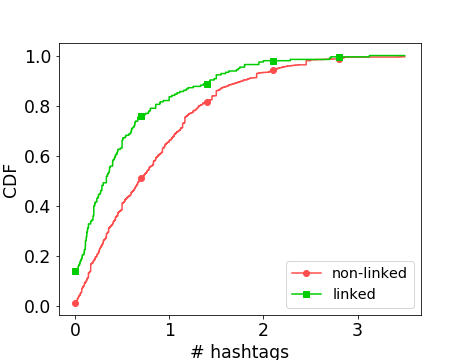}
			\caption{Hashtags.}
			\label{fig:ecdf-en-hashtags}
	\end{subfigure}
	
	\begin{subfigure}[b]{0.23\textwidth}
			\captionsetup{font=scriptsize}
			\includegraphics[width=\textwidth]{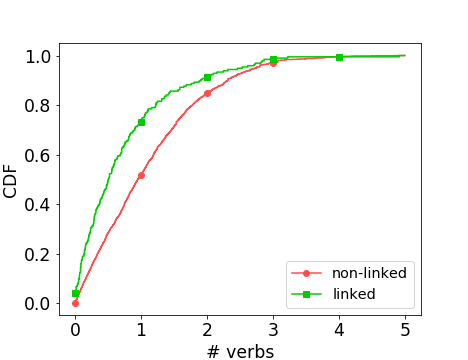}
			\caption{Verbs.}
			\label{fig:ecdf-en-verbs}
	\end{subfigure}
	\begin{subfigure}[b]{0.23\textwidth}
			\captionsetup{font=scriptsize}
			\includegraphics[width=\textwidth]{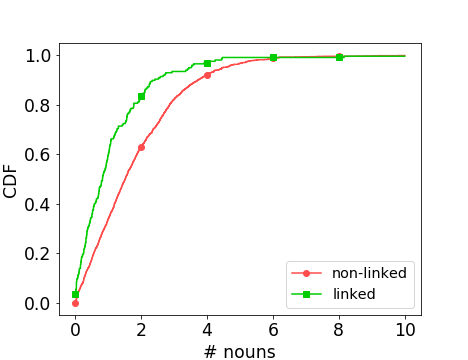}
			\caption{Nouns.}
			\label{fig:ecdf-en-nouns}
	\end{subfigure}
	
	\begin{subfigure}[b]{0.23\textwidth}
			\captionsetup{font=scriptsize}
			\includegraphics[width=\textwidth]{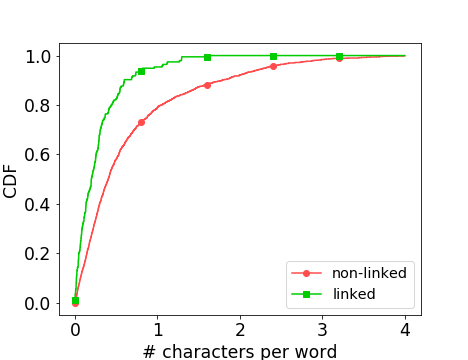}
			\caption{Mean \# characters per word.}
			\label{fig:ecdf-en-chars_per_word}
	\end{subfigure}
	\begin{subfigure}[b]{0.23\textwidth}
			\captionsetup{font=scriptsize}
			\includegraphics[width=\textwidth]{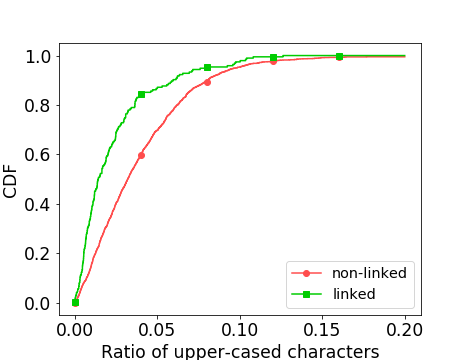}
			\caption{Upper-cased characters.}
			\label{fig:ecdf-en-uppercase}
	\end{subfigure}
	
	\caption{ECDF of (a) Mentions, (b) Hashtags, (c) Verbs, (d) Nouns, (e) Mean \# characters per word, and (f) Upper-cased characters.}
	\label{fig:ecdf-en}
\end{figure}

\descr{Activity Features.}
Figures~\ref{fig:ecdf-en-mentions}-\ref{fig:ecdf-en-hashtags} plot the ECDF for the number of mentions and hashtags for the linked and non-linked users ($p$$<$$0.01$). 
We observe that the non-linked users tend to have a higher difference in relation to the number of mentions and hashtags compared to the linked user accounts.
As for the inter-arrival time between the posted tweets (not shown in the plots), the difference is also statistically significant ($D$$=$$0.15849$).

\descr{Network Features.}
Table~\ref{tbl:used_features} presents the estimated network-based features.
To calculate such features, as already mentioned in Section~\ref{subsec:featuresextraction}, we consider the \textit{conversation-based network} constructed based on the mentions, replies, and retweets between each pair of users.
For the hub and authority scores the difference in distributions is statistically significant ($D$$=$$0.46745$) with mean (STD) values for the hub score to be equal to $0.0248$ ($0.0205$) and $0.0061$ ($0.0087$) for the linked and non-linked accounts, respectively, and for the authority score to be equal to $0.0238$ ($0.0196$) and $0.0058$ ($0.0083$) for the linked and non-linked accounts, respectively.
Concerning the pagerank and eigenvector centrality measures the difference is statistically significant ($D$$=$$0.49974$, $D$$=$$0.43939$, respectively), which is not the case for the clustering coefficient and the number of triangles where we cannot reject the null hypothesis that the distributions are different.

\descr{Linguistic Features.}
To identify the linkage of two or more accounts we consider a set of various linguistic attributes extracted from the available textual material.
Driven by the author profiling and identification tasks, we assume that the writing style of an author is unique enough to be distinguishable from the style of other authors~\cite{soler2017relevance}. 
In the literature for author profiling and identification a wide range of features is utilized; for instance, Burger et. al~\cite{burger2011discriminating} use more than $15M$ attributes, while Mukherjee and Liu more than $1K$~\cite{mukherjee2010improving}.
For our purposes, a more limited number of linguistic features is exploited, which has been shown to perform well in similar tasks ~\cite{soler2017relevance}.
This set of linguistic features is generic enough to capture the complexity and style of the discourse across different language families.
Indicatively, Figures~\ref{fig:ecdf-en-verbs}-\ref{fig:ecdf-en-uppercase} depict the ECDFs for the frequency of verbs, nouns, mean number of characters per word, and upper-cased characters features.
Comparing the distributions among the linked and non-linked accounts, we observe that the differences are statistically significant with $D$$=$$0.25181$, $D$$=$$0.29595$, $D$$=$$00.30405$, and $D$$=$$0.29209$, respectively.
Overall, in an effort to detect the linkage among users with the maximum possible efficiency we consider all the linguistic features presented in Table~\ref{tbl:used_features} (the difference in their distributions is statistically significant).

\begin{table}[!t]
\centering
\caption{Features evaluation.}
\label{tbl:features_evaluation}
\scalebox{0.75}{
\begin{tabular}{l|l}
\hline
\textbf{Dataset}                                                     & \textbf{Feature (preserving order)}                           \\ \hline\hline
\begin{tabular}[c]{@{}l@{}}Abusive\\ (English)\end{tabular}          & \begin{tabular}[c]{@{}l@{}}eigenvector (30\%), authority (10.29\%), hub (10.26\%)\\ pagerank (9.55\%), periods (6.83\%), stopwords (5.28\%) \\ diff. between longest - shortest words (4.98\%)\\ adverbial modifier (4.98\%), passive nominal subject (4.68\%)\\ mentions (4.51\%), coordination (4.43\%), adverbs (POS) (4.21\%)\end{tabular} \\ \hline
\begin{tabular}[c]{@{}l@{}}Terrorism\\ (Arabic)\end{tabular} & \begin{tabular}[c]{@{}l@{}}eigenvector (26.23\%), hashtags (8.45\%), punctuation (7.98\%)\\ mentions (7.78\%), diff. between longest - shortest words (7.18\%)\\ periods (7.03\%), adposition (6.93\%), mean\_max\_depth (6.46\%)\\ STD of word length (5.62\%), pagerank (5.51\%) \\ Hub (5.42\%), Authority (5.42\%)\end{tabular}            \\ \hline
\end{tabular}}
\end{table}

\descr{Note.}
The analysis presented  thus far was conducted on the English (abusive) dataset.
A similar analysis was conducted for the Arabic (terrorism-related) dataset; we omit the results due to space limits.

\descr{Features Evaluation.}
Table~\ref{tbl:features_evaluation} shows the top $12$ features for both the abusive and terrorism datasets based on the information gain approach which ranks features based on the information gain entropy in decreasing order.
We observe that in both cases the network features, which describe the connectivity of users in the network, are among the most contributing ones.
Especially for the abusive dataset such features seems to occupy the first places.
Regarding the activity features the \textit{average number of mentions} is among the top contributing ones in both cases, where especially for the terrorism-related dataset both the \textit{average number of hashtags} and \textit{mentions} seem to have a better discriminative ability comparing to the rest.
Focusing on the abusive dataset and the linguistic features, we observe that four out of seven are syntactic-based which indicates the importance of such features in distinguishing between linked and non-linked accounts.
Specifically, the most contributing syntactic-based features are the following: \textit{adverbs} (part-of-speech), \textit{adverbial modifier} (adverb or adverbial phrase that serves to modify a predicate or a modifier word), \textit{passive nominal subject} (a noun phrase which is the syntactic subject of a passive clause), and \textit{coordination} (is the relation between an element of a conjunct and the coordinating conjunction word of the conjunct).
With respect to the terrorism dataset and the linguistic features, we observe that the character-, word-, and syntactic-based ones tend to have an important discriminating power with the \textit{average number of punctuations} and the \textit{difference between the longest and shortest words} features being among the most contributing ones.

Overall, for the English (abusive) dataset, most of the features presented in Table~\ref{tbl:used_features} are useful (statistically  significant) in discriminating between the two classes (i.e., linked and non-linked user accounts).
However, some are not useful and are excluded to avoid adding noise.
Specifically, two features are excluded: the number of triangles and the clustering coefficient.
For the Arabic dataset, all features are useful and thus are used during the modeling analysis.

\subsection{Experimental Methodology}\label{subsec:expMethodology}

The features from the three categories $\{A, L, N\}$ that are selected as described above are employed for user modeling, while user pairs are modeled
based both on the absolute difference (\textit{abs}) and on the similarity of feature vectors (\textit{sim}); similarity is estimated based on Cosine similarity, and Euclidean and Manhattan distances. 
Therefore, the following approaches are evaluated: 
$Activity_{abs}$, 
$Liguistic_{abs}$, 
$Network_{abs}$,  
$All_{abs}$, and $All_{sim}$.
Moreover, the concatenation of $All_{abs}$ and $All_{sim}$ is also considered. 
In addition, the two features derived by modeling each pair of users using the edit distance and semantic similarities (see Section~\ref{subsec:usermodeling})
are considered in conjunction with the above, resulting in five additional approaches (see Table~\ref{tbl:classificationResultsAbusive}). 
Overall, a total of 11 different methods are evaluated.

We examined various machine learning algorithms, either probabilistic, tree-based, or ensemble classifiers, as well as deep neural networks.
For each family of classifiers, we only present those that achieve the best results (due to space limits).
Specifically, BayesNet, J48, and Random Forest (RF) are used as probabilistic, tree-based, and ensemble classifiers, respectively, along with the neural network setup.
We use WEKA for the traditional classifiers, and Keras with Theano~\cite{theano} for the deep learning models.
In all cases, we use repeated ($5$ times) $10$-fold cross validation which is less variable than the ordinary $10$-fold cross validation~\cite{kim2009estimating}.

\descr{Baseline.} 
Among the $11$ approaches, the first three (i.e., $Activity_{abs}$, $Linguistic_{abs}$, $Network_{abs}$) are our baselines.
Our aim is to not only determine the most effective classification approach, but to also assess whether the consideration of further information in the classification model (i.e., the features combined under different schemes) improves the overall performance, regardless of the choice of the classification algorithm.
As shown in Table~\ref{tbl:relatedAlternatives}, a wide range of activity, linguistic, and network features have been exploited in previous related research.
In an effort to be in alignment and comparable to literature to the maximum extent possible, here, we consider an important number of these features.
Specifically, we focus to those that are more applicable to our problem setting, since due to the inherent differences in  the structure of the various social media platforms, different features are applicable to each case.

At the same time, we further expand these features to better describe online user behavior.
Specifically, as for the linguistic features, we consider both dependency and tree features in addition to other commonly used ones (e.g., part-of-speech).
Moreover, a wider range of network features is extracted by building on top of the conversation-based network constructed using mentions, replies, and retweets; previous work has used only a reply-based network and considered only two network features.
Finally, to further improve the detection process, we also experiment with different combinations of features and user modeling approaches (i.e., absolute difference and similarity of feature vectors), while at the same time we further enhance the baseline by employing similarity-based features (i.e., edit distance and semantic similarity), which can encapsulate the authors' writing style in greater depth.

\descr{Evaluation metrics.} 
To be in alignment with similar works, standard evaluation metrics are reported: precision (prec), recall (rec), weighted area under the ROC curve (AUC), and accuracy (Acc).
In each table and for each evaluation metric (i.e., accuracy, AUC, precision, and recall), we highlight the top in terms of performance.

\begin{table*}[!t]
\caption{Classification results of BayesNet, J48, Random Forest, and Neural Network (Abusive Case, $X$$=$$200$).}
\scalebox{0.8}{
\begin{tabular}{lcccc|cccc|cccc|cccc}
                                                               & \multicolumn{4}{c|}{\textbf{BayesNet}}                            & \multicolumn{4}{c|}{\textbf{J48}}                                                                                                           & \multicolumn{4}{c|}{\textbf{Random Forest}}                        & \multicolumn{4}{c}{\textbf{Neural Network}}                                                                                               \\ \cline{2-17} 
                                                               & \textbf{Acc}   & \textbf{AUC}   & \textbf{Prec}  & \textbf{Rec}   & \multicolumn{1}{l}{\textbf{Acc}} & \multicolumn{1}{l}{\textbf{AUC}} & \multicolumn{1}{l}{\textbf{Prec}} & \multicolumn{1}{l|}{\textbf{Rec}} & \textbf{Acc}   & \textbf{AUC}   & \textbf{Prec}  & \textbf{Rec}   & \multicolumn{1}{l}{\textbf{Acc}} & \multicolumn{1}{l}{\textbf{AUC}} & \multicolumn{1}{l}{\textbf{Prec}} & \multicolumn{1}{l}{\textbf{Rec}} \\ \hline
\multicolumn{1}{l|}{Baseline $Activity_{abs}$}                         & 91.26          & 74.20          & 88.94          & 91.28          & 91.24                            & 59.34                            & 88.70                             & 91.14                             & 91.02          & 74.68          & 88.56          & 91.02          & 90.90                            & 65.00                            & 83.00                             & 91.00                            \\
\multicolumn{1}{l|}{Baseline $Linguistic_{abs}$}                       & 93.25          & 95.60          & 93.96          & 93.24          & 93.19                            & 79.88                            & 92.86                             & 93.18                             & 94.02          & 98.44          & 94.40          & 94.08          & 94.77                            & 91.65                            & 94.00                             & 95.00                            \\
\multicolumn{1}{l|}{Baseline $Network_{abs}$}                          & 97.60          & 96.78          & \textbf{97.58} & 97.60          & \textbf{99.08}                   & 94.92                            & \textbf{99.08}                    & \textbf{99.10}                    & 97.80          & 98.48          & 97.80          & 97.80          & 90.86                            & 81.41                            & 83.00                             & 91.00                            \\ \hline
\multicolumn{1}{l|}{$All_{abs}$}                              & 95.20          & \textbf{98.22} & 95.80          & 95.22          & 97.17                            & 89.40                            & 97.02                             & 97.18                             & 95.11          & 99.30          & 95.38          & 95.10          & 95.90                            & 96.05                            & \textbf{96.00}                    & \textbf{96.00}                   \\ \hline
\multicolumn{1}{l|}{$Activity_{abs}$ + $edits$ + $sem$}             & 97.02          & 97.52          & 96.96          & 97.02          & 98.77                            & 95.00                            & 98.78                             & 98.78                             & 97.43          & 98.76          & 97.44          & 97.44          & 90.90                            & 68.77                            & 83.00                             & 91.00                            \\
\multicolumn{1}{l|}{$Linguistic_{abs}$ + $edits$ + $sem$}           & 92.97          & 95.68          & 93.96          & 92.98          & 93.86                            & 80.98                            & 93.64                             & 93.86                             & 94.12          & 98.72          & 94.48          & 94.14          & 94.22                            & 90.92                            & 94.00                             & 94.00                            \\
\multicolumn{1}{l|}{$Network_{abs}$ + $edits$ + $sem$}            & \textbf{97.64} & 97.44          & \textbf{97.58} & \textbf{97.64} & 98.75                            & \textbf{95.30}                   & 98.74                             & 98.74                             & \textbf{97.81} & \textbf{99.50} & \textbf{97.82} & \textbf{97.82} & 90.86                            & 80.90                            & 83.00                             & 91.00                            \\ \hline
\multicolumn{1}{l|}{$All_{abs}$ + $edits$ + $sem$}                & 95.06          & \textbf{98.22} & 95.80          & 95.06          & 96.67                            & 87.50                            & 96.54                             & 96.68                             & 95.13          & 99.00          & 95.38          & 95.12          & \textbf{95.95}                   & 95.91                            & \textbf{96.00}                    & \textbf{96.00}                   \\ \hline
\multicolumn{1}{l|}{$All_{sim}$}                              & 88.61          & 86.38          & 90.14          & 88.60          & 93.94                            & 68.98                            & 93.96                             & 93.94                             & 94.25          & 90.28          & 94.10          & 94.26          & 91.95                            & 80.22                            & 93.00                             & 92.00                            \\
\multicolumn{1}{l|}{$All_{sim}$ + $All_{abs}$}               & 94.27          & 97.86          & 95.20          & 94.26          & 96.40                            & 88.04                            & 96.26                             & 96.40                             & 95.06          & 99.24          & 95.28          & 95.06          & 95.45                            & \textbf{96.13}                   & 95.00                             & 95.00                            \\
\multicolumn{1}{l|}{$All_{sim}$ + $All_{abs}$ + $edits$ + $sem$} & 94.05          & 97.86          & 95.22          & 94.06          & 96.50                            & 87.72                            & 96.32                             & 96.50                             & 95.02          & 99.38          & 95.28          & 95.04          & 95.45                            & 95.99                            & 95.00                             & 95.00                            \\ \hline
\end{tabular}}
\label{tbl:classificationResultsAbusive} 
\end{table*}

\section{Results}\label{sec:classificationResults}

\begin{table*}[!t]
\caption{Classification results of BayesNet, J48, Random Forest, and Neural Network (Terrorism Case, $X$$=$$200$).}
        \scalebox{0.8}{
\begin{tabular}{lcccc|cccc|cccc|cccc}
                                                               & \multicolumn{4}{c|}{\textbf{BayesNet}}                            & \multicolumn{4}{c|}{\textbf{J48}}                                                                                                           & \multicolumn{4}{c}{\textbf{Random Forest}}                        & \multicolumn{4}{c}{\textbf{Neural Network}}                                                                                               \\ \cline{2-17} 
                                                               & \textbf{Acc}   & \textbf{AUC}   & \textbf{Prec}  & \textbf{Rec}   & \multicolumn{1}{l}{\textbf{Acc}} & \multicolumn{1}{l}{\textbf{AUC}} & \multicolumn{1}{l}{\textbf{Prec}} & \multicolumn{1}{l|}{\textbf{Rec}} & \textbf{Acc}   & \textbf{AUC}   & \textbf{Prec}  & \textbf{Rec}   & \multicolumn{1}{l}{\textbf{Acc}} & \multicolumn{1}{l}{\textbf{AUC}} & \multicolumn{1}{l}{\textbf{Prec}} & \multicolumn{1}{l}{\textbf{Rec}} \\ \hline
\multicolumn{1}{l|}{Baseline $Activity_{abs}$}                         & 92.00          & 81.20          & 90.74          & 91.60          & 91.70                            & 77.22                            & 90.68                             & 91.70                             & 89.59          & 81.14          & 87.66          & 89.58          & 90.90                            & 78.23                            & 83.00                             & 91.00                            \\
\multicolumn{1}{l|}{Baseline $Linguistic_{abs}$}                       & 94.72          & 97.34          & 95.42          & 94.72          & 95.64                            & 84.62                            & 95.50                             & 95.64                             & 96.13          & 98.72          & 96.20          & 96.14          & 96.09                            & 96.40                            & 96.00                             & 96.00                            \\
\multicolumn{1}{l|}{Baseline $Network_{abs}$}                          & 87.95          & 95.86          & 93.84          & 87.96          & 95.69                            & 92.76                            & 96.38                             & 96.28                             & 96.02          & 94.28          & 96.08          & 96.04          & 91.00                            & 83.26                            & 92.00                             & 91.00                            \\ \hline
\multicolumn{1}{l|}{$All_{abs}$}                              & 96.60          & 99.18          & 97.08          & 96.60          & 96.73                            & 90.32                            & 96.62                             & 96.74                             & 97.06          & 99.38          & 97.02          & 97.00          & 96.18                            & 97.99                            & 96.00                             & 96.00                            \\ \hline
\multicolumn{1}{l|}{$Activity_{abs}$ + $edits$ + $sem$}             & 90.09          & 94.82          & 93.00          & 90.10          & 96.19                            & 80.18                            & 96.08                             & 96.20                             & 95.77          & 94.80          & 95.52          & 95.78          & 93.22                            & 93.52                            & 93.00                             & 93.00                            \\
\multicolumn{1}{l|}{$Linguistic_{abs}$ + $edits$ + $sem$}           & 94.73          & 97.74          & 95.60          & 94.74          & 96.85                            & 88.50                            & 96.74                             & 96.86                             & 96.63          & 99.00          & 96.62          & 96.62          & 96.68                            & 97.24                            & \textbf{97.00}                    & \textbf{97.00}                   \\
\multicolumn{1}{l|}{$Network_{abs}$ + $edits$ + $sem$}            & 95.22          & 99.22          & 96.52          & 95.22          & \textbf{97.71}                   & \textbf{94.16}                   & \textbf{97.68}                    & \textbf{97.70}                    & \textbf{97.56} & 98.84          & \textbf{97.48} & \textbf{97.56} & 94.54                            & 92.12                            & 94.00                             & 95.00                            \\
\hline
\multicolumn{1}{l|}{$All_{abs}$ + $edits$ + $sem$}                & 96.70          & \textbf{99.30} & \textbf{97.26} & 96.70          & 97.57                            & 93.28                            & 97.58                             & 97.58                             & 97.10          & \textbf{99.50} & 97.10          & 97.10          & 96.59                            & \textbf{98.45}                   & \textbf{97.00}                    & \textbf{97.00}                   \\ \hline
\multicolumn{1}{l|}{$All_{sim}$}                              & 94.92          & 94.10          & 94.94          & 94.92          & 94.66                            & 85.60                            & 94.38                             & 94.66                             & 95.66          & 94.58          & 95.40          & 95.66          & 93.63                            & 89.95                            & 93.00                             & 94.00                            \\
\multicolumn{1}{l|}{$All_{sim}$ + $All_{abs}$}               & \textbf{96.78} & 98.96          & \textbf{97.26} & \textbf{96.78} & 95.44                            & 86.88                            & 95.30                             & 95.46                             & 96.86          & 99.24          & 96.88          & 96.88          & 96.40                            & 97.59                            & 96.00                             & 96.00                            \\
\multicolumn{1}{l|}{$All_{sim}$ + $All_{abs}$ + $edits$ + $sem$} & 96.72          & 99.04          & \textbf{97.26} & 96.70          & 97.11                            & 91.58                            & 97.04                             & 97.10                             & 97.04          & 99.36          & 97.04          & 97.04          & \textbf{96.77}                   & 98.37                            & \textbf{97.00}                    & \textbf{97.00}                   \\ \hline
\end{tabular}}
\label{tbl:classificationResultsTerrorism}
\end{table*}

We first evaluate user identity linkage detection on the abusive dataset and then on the terrorism dataset.
The results are first presented on datasets built for $X$$=$$200$ and $Z$$=$$1,800$, and then for varying $X$ and $Z$ values.
Moreover, the presented results are based on randomly assigning tweets between linked accounts when building the ground truth; we achieve similar performance with interleaving (we omit these results due to space limits).

\subsection{Abusive Dataset  (English tweets)}\label{sec:abusivedataset}
Table~\ref{tbl:classificationResultsAbusive} shows that BayesNet achieves the best results when using the absolute difference for the user modeling, with AUC between \textbf{74.20\%} and \textbf{98.22\%} and accuracy between \textbf{91.26\%} and \textbf{97.64\%}.
We achieve the best precision and recall with the network features, either on their own (i.e., \textbf{97.58\%} and \textbf{97.60\%}) or combined with the two texts' similarity measures, i.e., edit distance (\textit{edits}) and semantic similarity (\textit{sem}), (i.e., \textbf{97.58\%} and \textbf{97.64\%}).
With regard to feature categories, the activity ones contribute the least, with \textbf{88.94\%} precision, \textbf{91.28\%} recall, and moderate AUC of \textbf{74.20\%}.

Similar to BayesNet, J48 achieves the best AUC (up to \textbf{95.30\%}) based on the absolute difference between features, while again we achieve the best performance using the network features (i.e., \textbf{99.08\%} precision and \textbf{99.10\%} recall).
Finally, texts' similarities appear to have an important role, since in most cases they tend to improve the classification results.

\begin{figure}[!t]
	\centering
	\begin{subfigure}[b]{0.5\textwidth}
			\includegraphics[width=\textwidth]{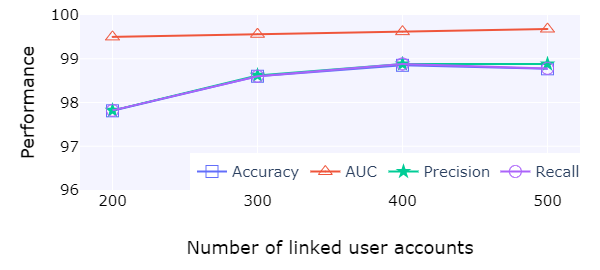}
			\caption{Linked instances.}
			\label{fig:en_varying.users}
	\end{subfigure}
	\begin{subfigure}[b]{0.49\textwidth}
			\includegraphics[width=\textwidth]{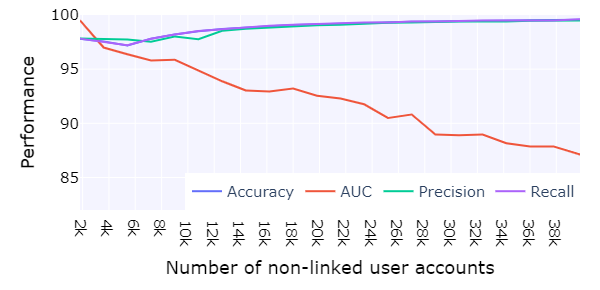}
			\caption{Non-linked instances.}
			\label{fig:en_dataset_unbalanced}
	\end{subfigure}
	\caption{Varied linked ($X$$=$$[$$200$$,$$500$$]$, step$=$$100$) and non-linked ($X$$=$$200$) instances.}
\end{figure}

\begin{figure}[!t]
	\centering
	\begin{subfigure}[b]{0.5\textwidth}
			\includegraphics[width=\textwidth]{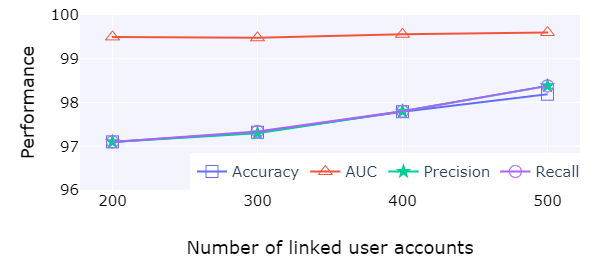}
			\subcaption{Linked instances.}
			\label{fig:ar_varying.users}
	\end{subfigure}
	\begin{subfigure}[b]{0.49\textwidth}
			\includegraphics[width=\textwidth]{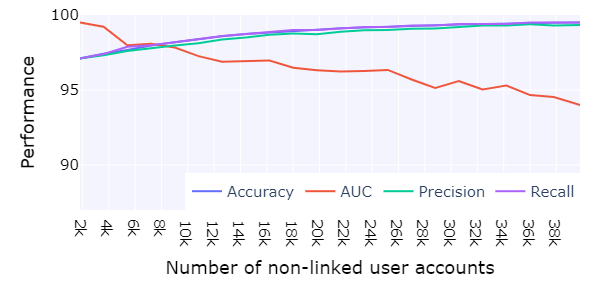}
			\subcaption{Non-linked instances.}
			\label{fig:en_dataset_unbalanced_ar}
	\end{subfigure}
	\caption{Varied linked ($X$$=$$[$$200$$,$$500$$]$, step$=$$100$) and non-linked ($X$$=$$200$) instances.}
\end{figure}

Contrary to the traditional classifiers, the linguistic features perform better in the NN setup (i.e., \textbf{91.65\%} AUC, \textbf{94\%} precision, and \textbf{95\%} recall) compared to the activity and network ones.
Overall, we obtain the best performance in terms of AUC (\textbf{96.13\%}) when all features are considered, when using both the absolute difference and similarity of features vectors for user modeling.
This indicates that the more information as input to the NN, the better the performance.

Finally, the Random Forest ensemble classifier achieves
the best performance when network features are used in addition to texts' similarities.
Specifically, AUC equals to \textbf{99.50\%} with precision, recall, and accuracy around \textbf{97.80\%}.
Compared to the probabilistic, tree-based classifiers, and deep neural networks, the Random Forest model achieves the best AUC, with precision and recall values  among the top; thus we use only this in the following experiments.

Thus far, we used the ground truth created with $X = 200$ randomly selected users.
Next, we vary $X$ from $200$ to $500$ with step $100$.
Figure~\ref{fig:en_varying.users}, which depicts the performance of the Random Forest model, shows that from $200$ to $300$ users there is a slight increase in precision, recall, and accuracy, while then the performance is quite stable with more than \textbf{99\%} AUC in all cases.

We also examine how the number of the non-linked instances (unbalanced dataset) affects the results.
The selected number of linked accounts equals to $200$, thus the upper limit of non-linked accounts equals to $39,800$.
Figure~\ref{fig:en_dataset_unbalanced} indicates that even with the highest number of non-linked user accounts, AUC remains at quite satisfactory levels (\textbf{87.30\%}).
Precision and recall increase as more data is available, while after a point ($\sim$$24k$ non-linked accounts) they are not significantly affected.
This is mainly attributed to the higher precision and recall values for the non-linked accounts.
Hence, even with a higher amount of non-linked accounts, 
the proposed model will succeed to effectively distinguish between linked and non-linked users.

\subsection{Terrorism Dataset (Arabic tweets)}

Table~\ref{tbl:classificationResultsTerrorism} shows that when using BayesNet, the linguistic features alone result in better performance compared to the activity and network ones.
We achieve the best precision (\textbf{97.26\%}) and recall (\textbf{96.78\%}) when we consider all feature categories together using both the absolute difference and the similarity of feature vectors for user modeling.
AUC maintains above \textbf{94\%}, for all cases, except when only the activity features are considered (\textbf{81.20\%} AUC).
Contrary to the BayesNet results in the abusive dataset, here we see that when the similarity of feature vectors (combined with additional features) is used as a user modeling method, we achieve high precision and recall values (up to \textbf{97.26\%} and \textbf{96.78\%}, respectively).

Out of the tree-based classifiers, J48 performs best (similar to the abusive case), following also a similar pattern in terms of the most well-performing feature categories and user modeling methods.
Network features appear to contribute more with the best performance (i.e., \textbf{97.68\%} precision, \textbf{97.70\%} recall, \textbf{94.16\%} AUC) achieved when combined with the texts' similarity measures.

Similar to the abusive case, linguistic features contribute more in the NN setup (\textbf{96.40\%} AUC, \textbf{96\%} precision and recall) compared to activity and network ones.
We obtain the best AUC (\textbf{98.45\%}) when all feature categories are considered, in addition to the texts' similarities.
In almost all cases, AUC, precision, and recall are higher than $90\%$, highlighting the stability of the used setup.

Finally, the best performance for the Random Forest (\textbf{99.50\%} AUC) is obtained when all features under the absolute difference modeling method are combined with the texts' similarities. 
Regarding the feature categories, linguistic features result in better performance compared to the rest (\textbf{98.72\%} AUC), which is also the case when combined with the texts' similarities (\textbf{99\%} AUC).
Overall, Random Forest leads to the best AUC and therefore is used next.

Figure~\ref{fig:ar_varying.users} shows the performance of Random Forest when the number of the selected linked accounts changes.
AUC is fairly stable with its value to be in all cases above \textbf{99\%}, which indicates the suitability of the proposed model.
Concerning the other evaluation metrics, the increase of the linked accounts results in higher values.
Figure~\ref{fig:en_dataset_unbalanced_ar} depicts how the proposed model performs with an unbalanced dataset (as in the abusive case: $200$ linked and up to $39,800$ non-linked accounts).
Overall, AUC fluctuates from \textbf{94\%} to \textbf{99.50\%}, which again points out the stability of the proposed model and precision and recall from \textbf{97.1\%} to $\approx$\textbf{99\%}.

\subsection{Classification Takeaways}
Overall, our models perform well for both
the abusive and terrorism-related 
datasets.
For instance, the high ROC area\footnote{AUC of the ROC curves are typically used to evaluate the performance (sensitivity) of a model.} for the overall classification
($99.50\%$ in both cases) indicates that the proposed models can quite successfully discriminate between linked and non-linked accounts.
Even though the performance is slightly different in terms of the precision, recall values and the classification models, in both studied cases
the traditional 
classifiers performed better.
The lower performance of the neural network model can be justified by the limited number of instances used for building the model, since NNs perform better when large numbers of training data is available.
Moreover, in most cases, a better performance is achieved when baseline features are enhanced with additional information.

Focusing on the specific feature categories, we observe that the network features contribute significantly to the classification (especially when traditional classifiers are used); this highlights the importance of considering the connectivity of a user in a network to detect more efficiently the linkage between users.

A quite important observation 
is that the proposed models perform well in different languages, and the performance, in some cases, is slightly better in the Arabic dataset.
This could possibly be attributed to the way that the initial data was collected.
The abusive dataset was created based on \#Gamergate as a seed word for querying Twitter, while then during the collection process further filtering keywords were added in consecutive time intervals to select additional abusive-related content~\cite{chatzakou2017measuring}.
On the contrary, the terrorism-related data was collected based on targeted filtering keywords from the very beginning.
Hence, the abusive dataset is less focused than the terrorist one, and thus users' behavioral patterns may differ more, making the classification somewhat harder.

Overall, even with more targeted or broader data, the proposed ensemble models succeed in distinguishing quite effectively between linked and non-linked accounts.
Moreover, we observe that, for both the abusive and terrorism datasets, the ensemble models built using the network features in addition to the texts’ similarity measures result in high performance (AUC > $98\%$ and Acc, Prec, Rec > $97\%$).
Hence, since some linguistic features are language-dependent and thus additional effort would be needed for constructing such models for other languages, one could opt for the network-based model which is easier to adapt to different languages (probably with a slight negative effect on the overall performance).

\section{Conclusions \& Future Work}\label{sec:conclusions}

Similar to the offline world, user-generated content in online social networks often relates to abusive or even illegal activities.
While social media administrators often take intensive actions to remove the content and respective content producers not complying with their rules, users with non-legitimate or abnormal activity often tend to create multiple accounts in an effort to bypass and to be a step ahead of the applied combating measures.
This work proposed a framework for detecting accounts likely to belong to the same natural person in an attempt to combat multiple non-legitimate accounts.
We considered several attributes of users' online activity, posts, and networks,
and traditional machine learning methods, as well as deep neural networks were tested.
The results showed that our method is able to effectively detect linked accounts 
related to non-legitimate, or even illegal (abusive and terrorism-related) activities, in different languages:
English and Arabic.

As future work, we plan to conduct our analysis on other online social media platforms, such as YouTube and Facebook, so as to understand if our methods can be easily adapted within and across other social networks.
Moreover, the proposed method could be extended to consider additional linguistic attributes, like sarcasm and irony.
Finally, we aim to also investigate the effectiveness of our framework in domains amenable to public opinion manipulation and propaganda, 
such as politics.

\begin{acks}
This research has received funding from the European Union's H2020 research and innovation programme as part of the CONNEXIONs (GA No 786731) and PREVISION (GA No 833115) projects.
\end{acks}

\bibliographystyle{abbrv}
\bibliography{bibliography}

\begin{thebibliography}{10}

\bibitem{babcock2014latent}
M.~J. Babcock, V.~P. Ta, and W.~Ickes.
\newblock Latent semantic similarity and language style matching in initial
  dyadic interactions.
\newblock {\em Journal of Language and Social Psychology}, 33(1):78--88, 2014.

\bibitem{badjatiya2017deep}
P.~Badjatiya, S.~Gupta, M.~Gupta, and V.~Varma.
\newblock Deep learning for hate speech detection in tweets.
\newblock In {\em Proceedings of the 26th International Conference on World
  Wide Web Companion}, pages 759--760. IW3C2, 2017.

\bibitem{burger2011discriminating}
J.~D. Burger, J.~Henderson, G.~Kim, and G.~Zarrella.
\newblock Discriminating gender on twitter.
\newblock In {\em Proceedings of the conference on empirical methods in natural
  language processing}, pages 1301--1309. Association for Computational
  Linguistics, 2011.

\bibitem{chatzakou2017mean}
D.~Chatzakou, N.~Kourtellis, J.~Blackburn, E.~De~Cristofaro, G.~Stringhini, and
  A.~Vakali.
\newblock Mean birds: Detecting aggression and bullying on twitter.
\newblock In {\em Proceedings of the 2017 ACM on Web Science Conference}, pages
  13--22. ACM, 2017.

\bibitem{chatzakou2017measuring}
D.~Chatzakou, N.~Kourtellis, J.~Blackburn, E.~De~Cristofaro, G.~Stringhini, and
  A.~Vakali.
\newblock Measuring \#gamergate: A tale of hate, sexism, and bullying.
\newblock In {\em Proceedings of the 26th International Conference on World
  Wide Web Companion}, pages 1285--1290. IW3C2, 2017.

\bibitem{conway2018disrupting}
M.~Conway, M.~Khawaja, S.~Lakhani, J.~Reffin, A.~Robertson, and D.~Weir.
\newblock Disrupting daesh: measuring takedown of online terrorist material and
  its impacts.
\newblock {\em Studies in Conflict \& Terrorism}, pages 1--20, 2018.

\bibitem{fisher2015swarmcast}
A.~Fisher.
\newblock Swarmcast: How jihadist networks maintain a persistent online
  presence.
\newblock {\em Perspectives on Terrorism}, 9(3):3--20, 2015.

\bibitem{founta2018unified}
A.-M. Founta, D.~Chatzakou, N.~Kourtellis, J.~Blackburn, A.~Vakali, and
  I.~Leontiadis.
\newblock A unified deep learning architecture for abuse detection.
\newblock In {\em Proceedings of the 2019 ACM on Web Science Conference (to
  appear)}. ACM, 2019.

\bibitem{gephi}
Gephi.
\newblock \url{https://gephi.org/}, {2019}.

\bibitem{gialampoukidis2017detection}
I.~Gialampoukidis, G.~Kalpakis, T.~Tsikrika, S.~Papadopoulos, S.~Vrochidis, and
  I.~Kompatsiaris.
\newblock Detection of terrorism-related twitter communities using centrality
  scores.
\newblock In {\em Proceedings of the 2nd International Workshop on Multimedia
  Forensics and Security}, pages 21--25. ACM, 2017.

\bibitem{goga2013exploiting}
O.~Goga, H.~Lei, S.~H.~K. Parthasarathi, G.~Friedland, R.~Sommer, and
  R.~Teixeira.
\newblock Exploiting innocuous activity for correlating users across sites.
\newblock In {\em Proceedings of the 22nd International Conference on World
  Wide Web}, pages 447--458. ACM, 2013.

\bibitem{goga2015reliability}
O.~Goga, P.~Loiseau, R.~Sommer, R.~Teixeira, and K.~P. Gummadi.
\newblock On the reliability of profile matching across large online social
  networks.
\newblock In {\em Proceedings of the 21th ACM SIGKDD International Conference
  on Knowledge Discovery and Data Mining}, pages 1799--1808. ACM, 2015.

\bibitem{hu2004mining}
M.~Hu and B.~Liu.
\newblock Mining and summarizing customer reviews.
\newblock In {\em Proceedings of the 10th ACM SIGKDD International Conference
  on Knowledge Discovery and Data Mining}, pages 168--177. ACM, 2004.

\bibitem{johansson2013detecting}
F.~Johansson, L.~Kaati, and A.~Shrestha.
\newblock Detecting multiple aliases in social media.
\newblock In {\em Proceedings of the 2013 IEEE/ACM international conference on
  advances in social networks analysis and mining}, pages 1004--1011. ACM,
  2013.

\bibitem{johansson2015timeprints}
F.~Johansson, L.~Kaati, and A.~Shrestha.
\newblock Timeprints for identifying social media users with multiple aliases.
\newblock {\em Security Informatics}, 4(1):7, 2015.

\bibitem{kayes2015ya-abuse}
I.~Kayes, N.~Kourtellis, D.~Quercia, A.~Iamnitchi, and F.~Bonchi.
\newblock The social world of content abusers in community question answering.
\newblock In {\em Proceedings of the 24th International Conference on World
  Wide Web}, pages 570--580. IW3C2, 2015.

\bibitem{kim2009estimating}
J.-H. Kim.
\newblock Estimating classification error rate: Repeated cross-validation,
  repeated hold-out and bootstrap.
\newblock {\em Computational statistics \& data analysis}, 53(11):3735--3745,
  2009.

\bibitem{klausen2015tweeting}
J.~Klausen.
\newblock Tweeting the jihad: Social media networks of {Western} foreign
  fighters in {Syria} and {Iraq}.
\newblock {\em Studies in Conflict \& Terrorism}, 38(1):1--22, 2015.

\bibitem{kumar2017army}
S.~Kumar, J.~Cheng, J.~Leskovec, and V.~Subrahmanian.
\newblock An army of me: Sockpuppets in online discussion communities.
\newblock In {\em Proceedings of the 26th International Conference on World
  Wide Web}, pages 857--866. IW3C2, 2017.

\bibitem{liu2016aligning}
L.~Liu, W.~K. Cheung, X.~Li, and L.~Liao.
\newblock Aligning users across social networks using network embedding.
\newblock In {\em Proceedings of the 25th International Joint Conference on
  Artificial Intelligence}, pages 1774--1780. AAAI Press, 2016.

\bibitem{liu2014hydra}
S.~Liu, S.~Wang, F.~Zhu, J.~Zhang, and R.~Krishnan.
\newblock Hydra: Large-scale social identity linkage via heterogeneous behavior
  modeling.
\newblock In {\em Proceedings of the 2014 International Conference on
  Management of data}, pages 51--62. ACM, 2014.

\bibitem{malhotra2012studying}
A.~Malhotra, L.~Totti, W.~Meira~Jr, P.~Kumaraguru, and V.~Almeida.
\newblock Studying user footprints in different online social networks.
\newblock In {\em Proceedings of the 2012 International Conference on Advances
  in Social Networks Analysis and Mining}, pages 1065--1070. IEEE Computer
  Society, 2012.

\bibitem{Massanari09102015}
A.~Massanari.
\newblock \#gamergate and the fappening: How reddit's algorithm, governance,
  and culture support toxic technocultures.
\newblock {\em New Media \& Society}, pages 329--346, 2017.

\bibitem{mcdonald2006multilingual}
R.~McDonald, K.~Lerman, and F.~Pereira.
\newblock Multilingual dependency analysis with a two-stage discriminative
  parser.
\newblock In {\em Proceedings of the 10th Conference on Computational Natural
  Language Learning}, pages 216--220. ACL, 2006.

\bibitem{MikolovWordEmbedding2013}
T.~Mikolov, K.~Chen, G.~Corrado, and J.~Dean.
\newblock Efficient estimation of word representations in vector space.
\newblock {\em CoRR}, abs/1301.3781, 2013.

\bibitem{mikolov2010recurrent}
T.~Mikolov, M.~Karafi{\'a}t, L.~Burget, J.~Cernock{\`y}, and S.~Khudanpur.
\newblock Recurrent neural network based language model.
\newblock In {\em Interspeech}, volume~2, page~3, 2010.

\bibitem{mu2016user}
X.~Mu, F.~Zhu, E.-P. Lim, J.~Xiao, J.~Wang, and Z.-H. Zhou.
\newblock User identity linkage by latent user space modelling.
\newblock In {\em Proceedings of the 22nd International Conference on Knowledge
  Discovery and Data Mining}, pages 1775--1784. ACM, 2016.

\bibitem{mukherjee2010improving}
A.~Mukherjee and B.~Liu.
\newblock Improving gender classification of blog authors.
\newblock In {\em Proceedings of the 2010 conference on Empirical Methods in
  natural Language Processing}, pages 207--217. Association for Computational
  Linguistics, 2010.

\bibitem{nakatanilangdetect}
S.~Nakatani.
\newblock Language detection library for java, 2010.

\bibitem{Navarro2001ApproximateStringMatching}
G.~Navarro.
\newblock A guided tour to approximate string matching.
\newblock {\em ACM Computing Surveys}, 33(1):31--88, 2001.

\bibitem{nie2016identifying}
Y.~Nie, Y.~Jia, S.~Li, X.~Zhu, A.~Li, and B.~Zhou.
\newblock Identifying users across social networks based on dynamic core
  interests.
\newblock {\em Neurocomputing}, 210:107--115, 2016.

\bibitem{pennekamp2019hi}
J.~Pennekamp, M.~Henze, O.~Hohlfeld, and A.~Panchenko.
\newblock Hi doppelg\"{a}nger : Towards detecting manipulation in news
  comments.
\newblock In {\em Companion Proceedings of The 2019 World Wide Web Conference},
  pages 197--205. ACM, 2019.

\bibitem{pennington2014glove}
J.~Pennington, R.~Socher, and C.~Manning.
\newblock Glove: Global vectors for word representation.
\newblock In {\em Proceedings of the 2014 Conference on Empirical Methods in
  Natural Language Processing (EMNLP)}, pages 1532--1543. ACL, 2014.

\bibitem{riederer2016linking}
C.~Riederer, Y.~Kim, A.~Chaintreau, N.~Korula, and S.~Lattanzi.
\newblock Linking users across domains with location data: Theory and
  validation.
\newblock In {\em Proceedings of the 25th International Conference on World
  Wide Web}, pages 707--719. IW3C2, 2016.

\bibitem{soler2017relevance}
J.~Soler-Company and L.~Wanner.
\newblock On the relevance of syntactic and discourse features for author
  profiling and identification.
\newblock In {\em Proceedings of the 15th Conference of the European Chapter of
  the Association for Computational Linguistics}, pages 681--687. ACL, 2017.

\bibitem{soliman2017aravec}
A.~B. Soliman, K.~Eissa, and S.~R. El-Beltagy.
\newblock Aravec: A set of arabic word embedding models for use in arabic nlp.
\newblock {\em Procedia Computer Science}, 117:256--265, 2017.

\bibitem{solorio2013case}
T.~Solorio, R.~Hasan, and M.~Mizan.
\newblock A case study of sockpuppet detection in wikipedia.
\newblock In {\em Proceedings of the Workshop on Language Analysis in Social
  Media}, pages 59--68. ACL, 2013.

\bibitem{twitterStats}
{Statista}.
\newblock {Number of monthly active Twitter users worldwide from 1st quarter
  2010 to 1st quarter 2019}.
\newblock \url{goo.gl/JLy8Ko}, 2019.

\bibitem{theano}
Theano.
\newblock \url{http://deeplearning.net/software/theano/}, {2019}.

\bibitem{tsikerdekis2014multiple}
M.~Tsikerdekis and S.~Zeadally.
\newblock Multiple account identity deception detection in social media using
  nonverbal behavior.
\newblock {\em IEEE Transactions on Information Forensics and Security},
  9(8):1311--1321, 2014.

\bibitem{twitterCompanySuspension}
{Twitter Public Policy}.
\newblock {Expanding and building \#TwitterTransparency}, April 2018.
\newblock \url{http://bit.ly/2SIHGNf}.

\bibitem{cnetfake}
Q.~Wong.
\newblock Facebook pulls down fake accounts from the uk and romania.
\newblock \url{cnet.co/2w81ZJd}, March 2019.

\bibitem{zafarani2013connecting}
R.~Zafarani and H.~Liu.
\newblock Connecting users across social media sites: a behavioral-modeling
  approach.
\newblock In {\em Proceedings of the 19th ACM SIGKDD International Conference
  on Knowledge Discovery and Data Mining}, pages 41--49. ACM, 2013.

\end{thebibliography}

\end{document}